# THE JOURNAL OF PHYSICAL CHEMISTRY C

Article

pubs.acs.org/JPCC

# Plasmonic Bubble Nucleation and Growth in Water: Effect of Dissolved Air

Xiaolai Li,[†,⊥] Yuliang Wang,[*,⊥,#] Mikhail E. Zaytsev,[†,‡] Guillaume Lajoinie,[†,§] Hai Le The,[†,∥] Johan G. Bomer,[∥] Jan C. T. Eijkel,[∥] Harold J. W. Zandvliet,[‡] Xuehua Zhang,[†,∇] and Detlef Lohse[*,†,○]

[†]Physics of Fluids, Max Planck Center Twente for Complex Fluid Dynamics and J.M. Burgers Centre for Fluid Mechanics, MESA+ Institute, [‡]Physics of Interfaces and Nanomaterials, MESA+ Institute, [§]TechMed Centre, and [∥]BIOS Lab-on-a-Chip Group, MESA+ Institute, University of Twente, P.O. Box 217, 7500AE Enschede, The Netherlands

[⊥]Robotics Institute, School of Mechanical Engineering and Automation and [#]Beijing Advanced Innovation Center for Biomedical Engineering, Beihang University, 37 Xueyuan Road, Haidian District, Beijing 100191, P.R. China

[∇]Department of Chemical and Materials Engineering, Donadeo Innovation Centre for Engineering, University of Alberta, Edmonton, Alberta T6G 1H9, Canada

[○]Max Planck Institute for Dynamics and Self-Organization, Am Fassberg 17, 37077 Göttingen, Germany

**ABSTRACT:** Under continuous laser irradiation, noble metal nanoparticles immersed in water can quickly heat up, leading to the nucleation of so-called plasmonic bubbles. In this work, we want to further understand the bubble nucleation and growth mechanism. In particular, we quantitatively study the effect of the amount of dissolved air on the bubble nucleation and growth dynamics, both for the initial giant bubble, which forms shortly after switching on the laser and is mainly composed of vapor, and for the final life phase of the bubble, during which it mainly contains air expelled from water. We found that the bubble nucleation temperature depends on the gas concentration: the higher the gas concentration, the lower the bubble nucleation temperature. Also, the long-term diffusion-dominated bubble growth is governed by the gas concentration. The radius of the bubbles grows as $R(t) \propto t^{1/3}$ for air-equilibrated and air-oversaturated water. In contrast, in partially degassed water, the growth is much slower since, even for the highest temperature we achieve, the water remains undersaturated.

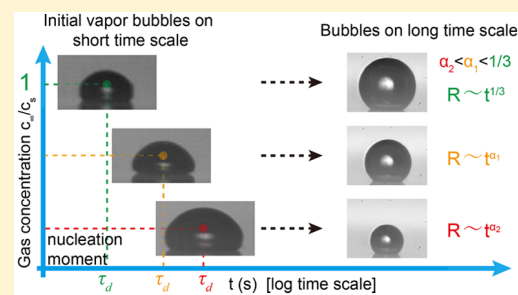

## ■ INTRODUCTION

When noble nanoparticles immersed in water are irradiated by a continuous wave laser at their resonance frequency, a huge amount of heat is explosively produced. This leads to a rapid temperature increase and hence to the vaporization of the surrounding water, resulting in the formation of microsized bubbles. These bubbles are referred to as plasmonic bubbles.[1−3] The plasmonic bubbles are relevant to various applications, such as biomedical diagnosis and cancer therapy,[4−6] solar energy harvesting,[2,7−9] micromanipulation of nano-objects,[10−13] and locally enhanced chemical reactions.[1,14] To exploit their potential applications, it is of key importance to understand the nucleation mechanism and explore the growth dynamics of these bubbles.

The formation of plasmonic bubbles involves complex physical processes, including optothermal conversion, heat transfer, phase transitions, gas diffusion, and many others.[2,3,15−18] In these processes, many factors are relevant, such as particle arrangement,[13] laser power,[19] liquid type,[20] and gas concentration.[3,15] Among these factors, gas concentration plays a crucial role in the formation and growth of plasmonic bubbles, like in the formation of other types of micro/nanoscale surface bubbles.[21−23] This has been ad-dressed in several studies[3,15,24] but ignored by many others.[25−27] By investigating the shrinkage dynamics of plasmonic bubbles, Baffou and co-workers found that the plasmonic bubbles could survive for several hundreds of seconds after the irradiation laser was switched off.[15] This is due to the fact that the bubbles are actually not vapor bubbles but mainly contain gas that was originally dissolved in the liquid.[17] Later, Liu et al. generated microbubbles at highly ordered plasmonic nanopillar arrays and observed a larger growth rate in air-equilibrated water than that in partially degassed water, which underlines the important role that the dissolved gas plays in the bubble formation.[3]

Recently, we conducted a systematic study of the plasmonic bubble nucleation mechanism[18] and growth dynamics.[17] We found that plasmonic microbubble nucleation and evolution in water can be divided into four phases: an initial giant vapor bubble (phase 1), oscillating bubbles (phase 2), a vaporization-dominated growth phase (phase 3), and, finally, a gas diffusion-dominated growth phase (phase 4).[17,18] Dissolved gas governs the bubble dynamics, especially for phases 1 and 4. In phase 1, water surrounding the laser-irradiated gold nanoparticles becomes superheated, that is, reaching a temperature that substantially exceeds the boiling temperature.[15,16,28−31] Consequently, vapor bubbles nucleate, grow, and then become microsized bubbles. In partially degassed water, the lower gas concentration leads to less nuclei and thus to suppression of the nucleation of bubbles. As a result, the superheat temperature in partially degassed water is higher than in air-equilibrated water, and it requires a longer illumination time in partially degassed water before the nucleation of the bubbles sets in. In phase 4, bubbles enter the long-term growth regime. This phase is dominated by the influx of dissolved gas from the surrounding water. Experimental results show that the bubble radius roughly scales as $R(t) \propto t^{1/3}$ in air-equilibrated water, which is distinctly different from the scaling in partially degassed water,[17] where the bubbles in the long-term hardly grow at all. For both kinds of behavior, $R(t) \propto t^{1/3}$ for the gaseous case and $R(t) \approx$ const for the strongly degassed case, a theoretical explanation was given.[18]

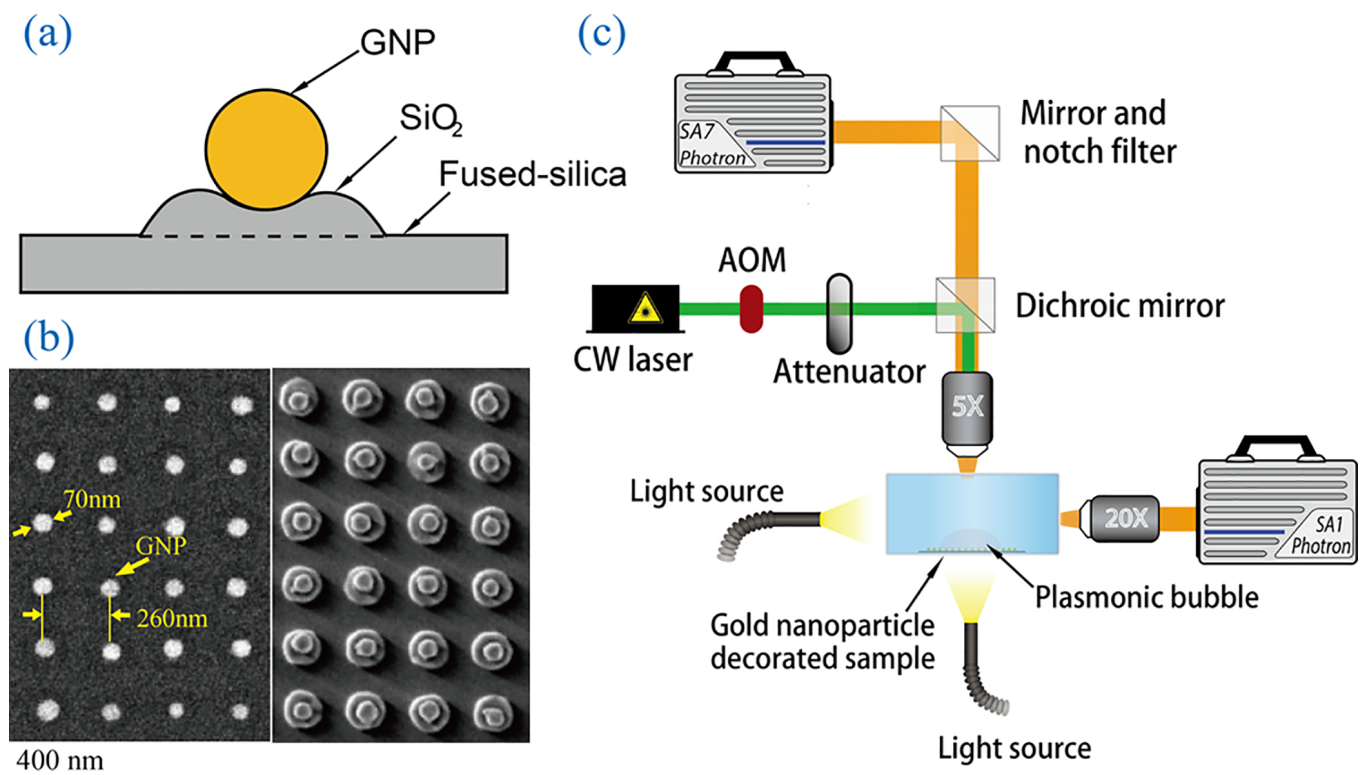

**Figure 1.** (a) Schematic of a gold nanoparticle sitting on a $SiO_2$ island on a fused-silica substrate, (b) energy-selective backscatter (left) and scanning electron microscopy (right) image of the patterned gold nanoparticle sample surface, and (c) schematic diagram of the optical setup for plasmonic microbubble imaging.

Despite the above-mentioned studies of the dissolved gas effect on bubble nucleation and growth dynamics, a quantitative understanding and investigation for systematically varying gas saturation are still lacking. In this work, such a systematic investigation is performed at six different relative concentrations $c_\infty/c_s$ of dissolved gas, ranging from highly degassed to oversaturated water. Here, $c_\infty$ is the actual gas concentration, and $c_s$ is the saturation concentration. This allows us to establish the detailed dependence of the dynamic parameters of bubble formation on the gas concentration. The understanding from this work facilitates the control of plasmonic bubble formation in related applications.

## ■ METHODS

**Sample Preparation.** A fused-silica surface patterned with an array of gold nanoparticles was used to produce plasmonic bubbles. A gold layer of ∼45 nm was first deposited on an amorphous fused-silica wafer by using an ion-beam sputtering system (home-built T′COathy machine, MESA+ NanoLab, Twente University). A bottom antireflection coating (BARC) layer (∼186 nm) and a photoresist (PR) layer (∼200 nm) were subsequently coated on the wafer. Periodic nanocolumns with diameters of ∼110 nm were patterned in the PR layer using displacement Talbot lithography (PhableR 100C, EULITHA).[32] These periodic PR nanocolumns were subsequently transferred at the wafer level to the underlying BARC layer, forming 110 nm BARC nanocolumns by using nitrogen plasma etching (home-built TEtske machine, Nano-Lab) at 10 mTorr and 25 W for 8 min. Using these BARC nanocolumns as a mask, the Au layer was subsequently etched by ion beam etching (Oxford i300, Oxford Instruments, U.K.) with 5 sccm Ar and 50−55 mA at an inclined angle of 5°. The etching for 9 min resulted in periodic Au nanodots supported on cone-shaped fused-silica features. The remaining BARC was stripped using oxygen plasma for 10 min (TePla 300E, PVA TePla AG, Germany). The fabricated array of Au nanodots was heated to 1100 °C in 90 min and subsequently cooled passively to room temperature. During the annealing process, these Au nanodots reformed into spherical-shaped Au nanoparticles. Figure 1a shows the schematic of a gold nanoparticle sitting on a $SiO_2$ island on a fused silica. The energy-selective backscatter (ESB) and scanning electron microscopy (SEM) images of the patterned gold nanoparticle sample surface are shown in Figure 1b, left and right, respectively.

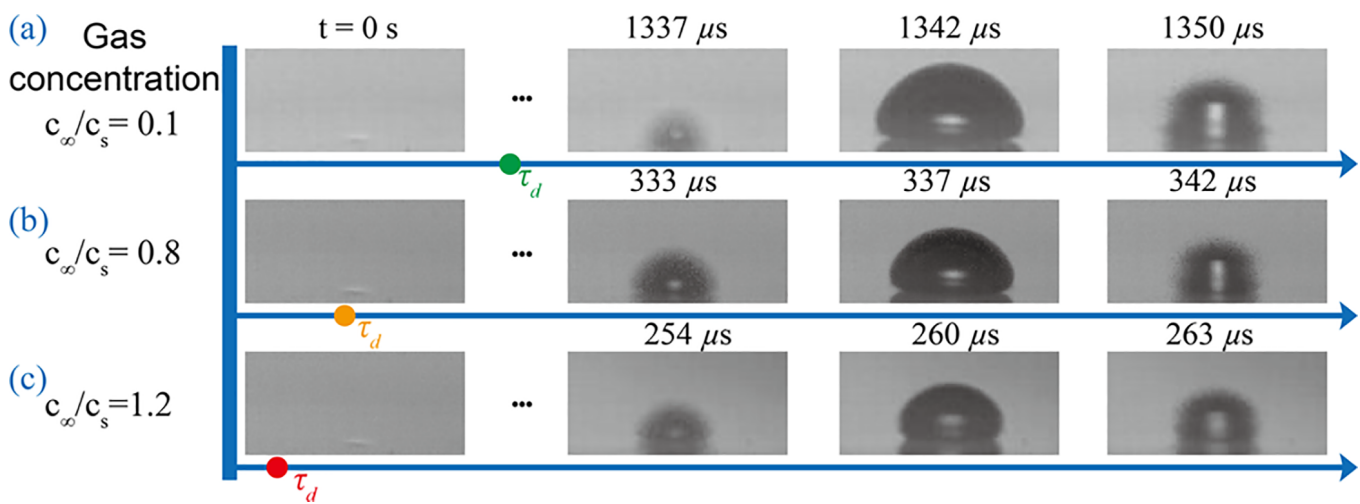

**Figure 2.** Initial giant bubble nucleation at different gas concentration levels of (a) $c_\infty/c_s = 0.1$, (b) $c_\infty/c_s = 0.8$, and (c) $c_\infty/c_s = 1.2$ at the same laser power $P_l = 130$ mW. The giant bubbles nucleate after a delay ($\tau_d$), which decreases with the increasing gas concentration, namely, $\tau_d(c_\infty/c_s = 1.2) < \tau_d(c_\infty/c_s = 0.8) < \tau_d(c_\infty/c_s = 0.1)$, while the maximum bubble sizes are smaller at higher gas concentrations.

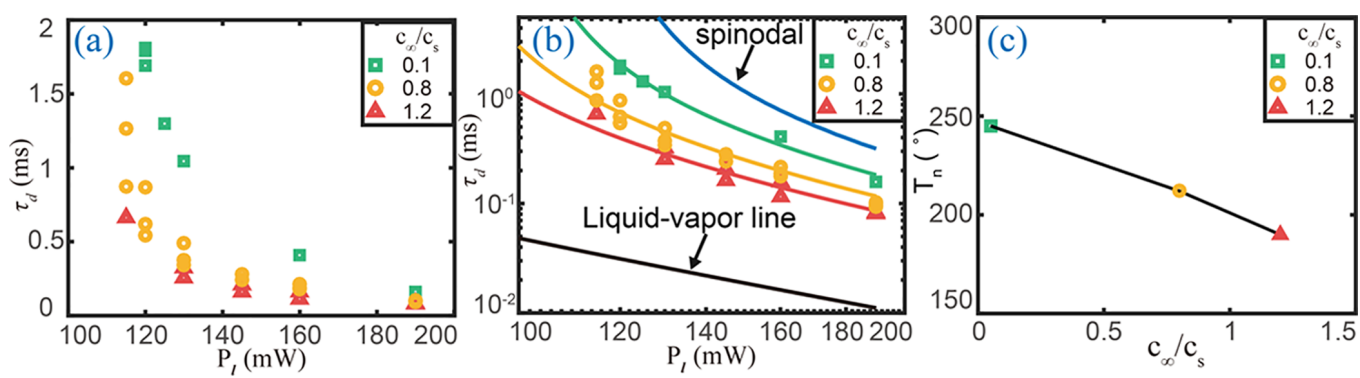

**Figure 3.** Dependence of the delay time $\tau_d$ on gas concentration for giant bubble nucleation. (a) Measured delay time $\tau_d$ as a function of laser power $P_l$ at different gas concentrations. At a given laser power, the higher the gas concentration, the shorter the delay time. (b) Double-logarithmic plot of $\tau_d$ versus $P_l$. The measured $\tau_d$ is fitted using a heat diffusion model, whose results are displayed with solid lines. The three curves are all located in between the boiling temperature curve and the spinodal curve. (c) Obtained nucleation temperature $T_n$ by fitting $\tau_d$ for gas concentration $c_\infty/c_s = 0.1$, 0.8, and 1.2. The higher gas concentration results in a lower nucleation temperature, which indicates that the dissolved gas facilitates the bubble nucleation.

**Gas Concentration Control.** During experiments, the nanoparticle-decorated sample surface was immersed in deionized (DI) water (Milli-Q Advantage A10 System, Germany). Experiments were separately executed to explore the effect of gas concentration both on the nucleation dynamics of the initial giant bubbles and on the long-term diffusive growth dynamics of the plasmonic bubbles. For the latter, six different gas concentration levels of $c_\infty/c_s = 0.55$, 0.65, 0.75, 0.85, 0.96, and 1.20 were used. For the initial giant bubbles, the nucleation dynamics only depends weakly on the gas concentration, and therefore, we have only considered three different gas concentrations of $c_\infty/c_s = 0.10$, 0.80, and 1.20. All experiments were conducted at 25 °C and 1 atm.

The oversaturated water was prepared by keeping the water at 4 °C for 12 h and then warming up to the room temperature of 25 °C. After that, the measured gas concentration $c_\infty/c_s$ was 1.2 (by an oxygen meter, Fibox 3 Trace, PreSens). For the air-equilibrated water, a sample bottle containing DI water was kept open in air for 10 h, and the measured gas concentration is 1.0. Similarly, the nearly air-equilibrated water with $c_\infty/c_s = 0.96$ was obtained by keeping the water in air for 8 h. To prepare partially degassed water with different gas concen-tration values, the water was first degassed for 30 min in a vacuum chamber, and then the concentration of 0.1 was obtained. Subsequently, we kept the bottle of the highly degassed water open in air, with the probe of the oxygen meter immersed in the water. The gas concentration in the water was adjusted by varying the air exposure time.

**Setup Description.** The experimental setup for plasmonic microbubble imaging is shown in Figure 1c. The gold nanoparticle-decorated sample was placed in a quartz glass cuvette and filled with water. A continuous-wave laser (Cobolt Samba) of 532 nm wavelength with a maximum power of 300 mW was used for sample irradiation. An acousto-optic modulator (Opto-Electronic, AOTFncVIS) was used as a shutter to control the laser irradiation on the sample surface. A pulse/delay generator (BNC model 565) was used to generate two different laser pulses of 400 $\mu$s and 4 s to study the short-term and long-term dynamics of microbubbles, respectively. The laser power was controlled by using a half-wave plate and a polarizer and measured by a photodiode power sensor (S130C, ThorLabs). Two high-speed cameras were installed in the setup. One (Photron SA7) was equipped with a 5× long working distance objective (LMPLFLN, Olympus), and the other (Photron SA1) was equipped with various long working distance objectives [10× (LMPLFLN, Olympus) and 20× (SLMPLN, Olympus)] and operated at various frame rates from 5 kfps up to 540 kfps. The first camera was used for top-view imaging, and the second one was for side-view imaging. Two light sources, Olympus ILP-1 and Schott ACE I, were applied to provide illumination for the two high-speed cameras. The optical images are processed with a home-designed image segmentation algorithm for the optimized extraction of the bubble radius in MATLAB.[33−35]

## RESULTS AND DISCUSSION

**Giant Bubble Nucleation.** The influence of gas concentration on the nucleation of plasmonic bubbles was investigated with a high-speed camera at a frame rate of 540 kfps. Figure 2 shows the nucleation and evolution of giant bubbles for three different gas concentration levels of $c_\infty/c_s$ = 0.1, 0.8, and 1.2, respectively, at the same laser power $P_l$ = 130 mW. The moment that the laser is switched on is taken as the origin of time, that is, $t$ = 0 s. One can see that, upon laser irradiation, the vapor microbubbles nucleate after a delay time $\tau_d$. Subsequently, the bubbles rapidly grow and reach a maximum size in several microseconds (~5 $\mu s$ after nucleation), followed by a sudden collapse. As can be seen in Figure 2, the maximum volume $V_{max}$ of the plasmonic bubbles counterintuitively decreases with increasing $c_\infty/c_s$, as discussed in ref 18. This can be explained by the increase in delay time $\tau_d$.[18] A longer delay time results in a larger amount of energy dumped into the system (before nucleation) and thus to a larger initial giant bubble.

The dependence of the delay time $\tau_d$ on the laser power $P_l$ at each gas concentration was also investigated, as shown in Figure 3a. Due to the limited spatial resolution (~1 $\mu m$) of the optical imaging system and the limited temporal resolution (frame rate, 540 kfps) of the high-speed camera, the estimated error in the determination of the delay time is ~2 $\mu s$. Since the delay time is in the order of a few hundreds of microseconds, the relative error is less than 1%. It is clear that, at each gas concentration $c_\infty/c_s$, the delay time decreases with increasing laser power $P_l$. For a given laser power value, the delay time decreases with an increasing gas concentration in water.

To determine the nucleation temperature for each combination of laser power $P_l$ and gas concentration $c_\infty/c_s$, we have solved a simple heat diffusion model (see ref 18). By assuming a spherical geometry and constant thermal properties, from that model, the temperature evolution around a nanoparticle can be computed by solving the thermal diffusion equation through a Fourier transformation. The thermal field generated by the nanoparticle array is estimated through a linear superposition of the temperature field generated by the individual nanoparticles within the Gaussian laser beam.[18] The numerical results from this approach were used to fit the measured delay time $\tau_d$ with a root-mean-square minimization method. The fitted curves for the three concentration cases are shown in the double-logarithmic plot of Figure 3b. In Figure 3b, the liquid−vapor line (i.e., $T_b$=100 °C for the boiling temperature) and spinodal line ($T_s$ = 305 °C for an ambient pressure of 1 atm) were determined by numerically calculating how long it takes to heat the liquid up to 100 and 305 °C, respectively, for given laser power $P_l$. The estimated nucleation temperatures from the fitting process are $T_1$ = 190 °C ($c_\infty/c_s$ = 0.1), $T_2$ = 212 °C ($c_\infty/c_s$ = 0.8), and $T_3$ = 245 °C ($c_\infty/c_s$ = 1.2). Figure 3c shows the decreasing nucleation temperature with increasing gas concentration level of $c_\infty/c_s$ from 0.1 to 1.2. This quantifies how the gas dissolved in water facilitates the bubble nucleation.

The dependence of the vapor bubble nucleation on dissolved gas has been pointed out in several other studies. Indeed, the presence of tiny cracks, cavities or pits filled with gas, and impurities or aggregations of gas molecules can act as nuclei for bubble nucleation.[15,16,28−31] These impurities in water reduce the nucleation temperature $T_n$ to a value substantially lower than the liquid spinodal temperature[36−38] and thus enhance the nucleation of bubbles. When the gas concentration is lower, the probability of forming gas nuclei is statistically reduced,[39] resulting in a higher $T_n$. Accordingly, the delay time $\tau_d$ will increase with decreasing gas concentration, as demonstrated in Figure 3.

In Figure 4, the maximum bubble size $V_{max}$ is plotted versus the total dumped energy $E = P_l\tau_d$, which is defined as the accumulated laser energy in the illumination spot on the substrate from the moment the laser is switched on until the moment of bubble nucleation. Here, the laser energy input during the period of bubble growth to its maximum volume is neglected. The reason that this can be done is that, once a giant bubble is formed on the substrate, it isolates the water from the gold nanoparticles, resulting in a strongly suppressed energy transfer to the water. Moreover, it normally only takes ~5 $\mu s$ for the initial giant bubble to grow to its maximum value right after it appears. This period of time (5 $\mu s$) is much smaller compared to the bubble nucleation delay time $\tau_d$ of a few hundreds of microseconds to milliseconds. As a result, the contribution of the energy transfer after bubble nucleation is very small and can safely be ignored.

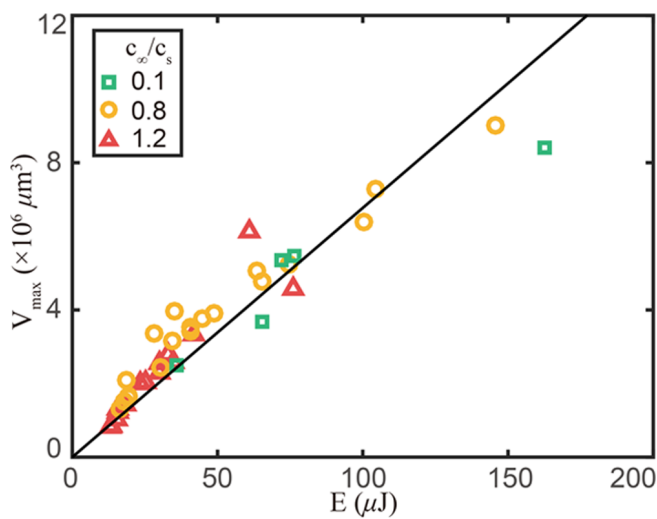

**Figure 4.** Maximum bubble volume $V_{max}$ versus deposited energy $E = P_l\tau_d$ at different gas concentration levels. For all gas concentrations, the same linear relation between $V_{max}$ and $E$ is found, regardless of the actual value of $\tau_d$ and $P_l$.

We find an universal linear relation between $V_{max}$ and $E$ for all gas concentrations. This universal linear relation $V_{max} = kE$ reflects that the energy stored in the vicinity of the nuclei determines how many water molecules can be vaporized. The maximum volume of the giant bubble only depends on the amount of energy dumped into the system before nucleation and not on the relative gas concentration. The latter confirms that the giant bubbles mainly consist of vapor.

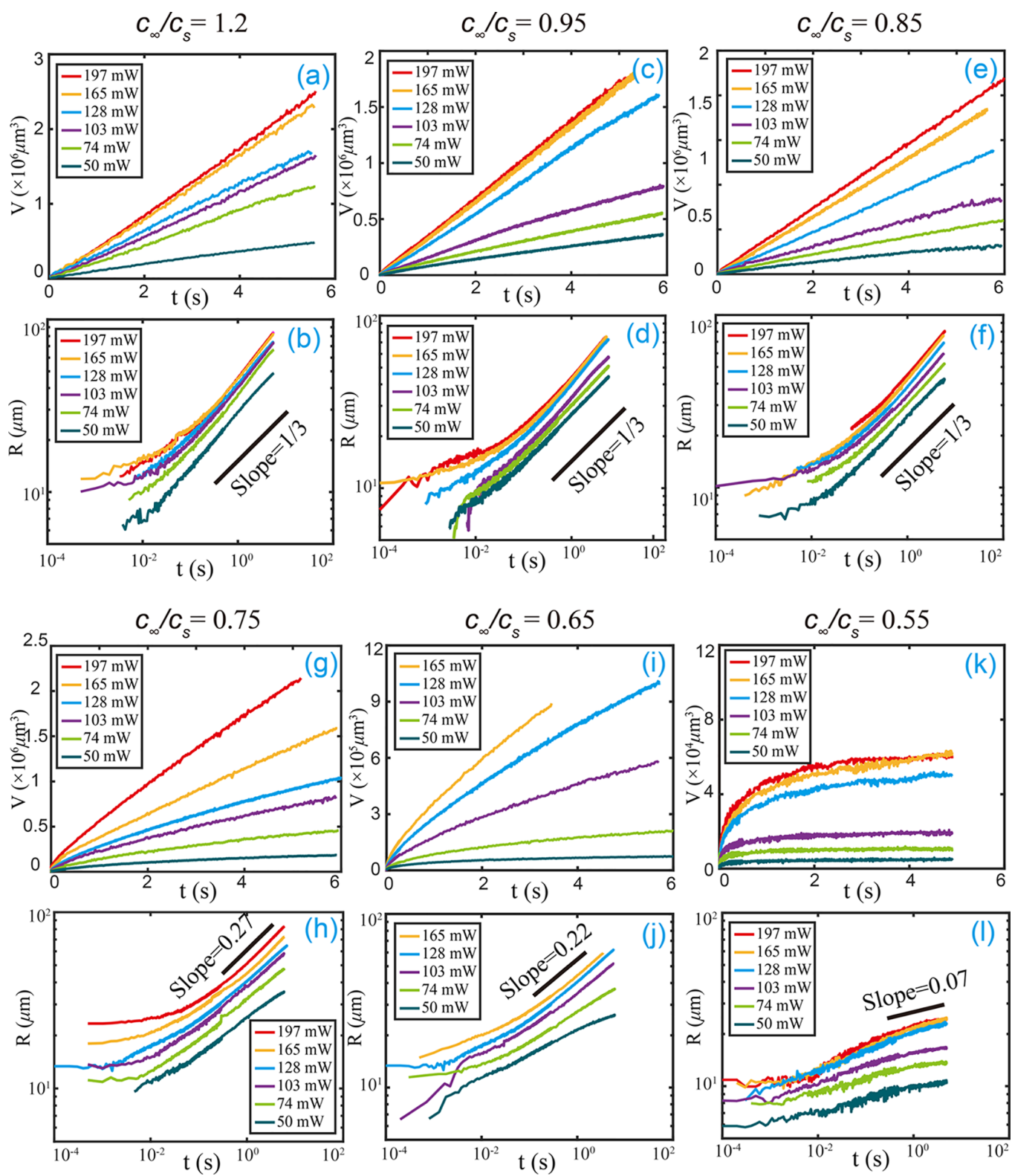

**Figure 5.** Long-term bubble growth dynamics at different gas concentrations and laser powers. (a, b) $c_\infty/c_s = 1.2$, (c, d) $c_\infty/c_s = 0.95$, (e, f) $c_\infty/c_s = 0.85$, (g, h) $c_\infty/c_s = 0.75$, (i, j) $c_\infty/c_s = 0.65$, and (k, l) $c_\infty/c_s = 0.55$. Linear plots of bubble volume $V$ (a, c, e, g, i, and k) and double-logarithmic plots of bubble radius $R$ (b, d, f, h, j, and l) as functions of time $t$ for gas concentrations $c_\infty/c_s = 1.2, 0.95, 0.85, 0.75, 0.65$, and 0.55, respectively. At $c_\infty/c_s = 1.2$ (oversaturated water), 0.95 (nearly air-equilibrated water), and 0.85 (partially degassed water), $V$ linearly increases with time for different laser powers. Accordingly, $R(t)$ follows the $t^{1/3}$ scaling law. At lower gas concentrations, the volume no longer linearly increases with $t$ but exhibits a reduced power law dependence of $R(t) \propto t^\alpha$ with $\alpha = 0.27, 0.22$, and 0.07 for $c_\infty/c_s = 0.75, 0.65$, and 0.55, respectively, as shown in (h), (j), and (l).

**Long-Term Growth Dynamics.** From the above analysis, we have obtained a better understanding of the gas dependence of the bubble nucleation dynamics. After the initial giant bubble collapses, it sequentially enters phase 2 (oscillating bubble phase) and phase 3 (vaporization-dominated growth phase), followed by the diffusive growth in phase 4.[18] Normally, the final phase begins at $t > 0.5$ s. We now focus on this long-term growth dynamics of the bubbles to investigate the role of the dissolved gas also on this terminal growth phase, again by varying the gas concentration levels from undersaturation to supersaturation (now $c_\infty/c_s = 0.55$–1.2).

Figure 5 shows the long-term growth dynamics of plasmonic bubbles at six concentration levels of $c_\infty/c_s = 1.2, 0.95, 0.85, 0.75, 0.65,$ and $0.55$. For the oversaturated water ($c_\infty/c_s = 1.2$), the bubble volume $V$ as a function of time is shown for several laser powers in Figure 5a. The bubble volume $V$ increases linearly with time $t$ for all laser powers. The corresponding bubble radius $R(t) \propto V(t)^{1/3}$ is shown in Figure 5b but now as a double-logarithmic plot. After $\sim 0.5$ s, $R$ roughly follows an effective $R(t) \sim t^{1/3}$ scaling law. Figure 5c,d shows the corresponding results in a nearly gas-equilibrated case with $c_\infty/c_s = 0.95$. The bubble growth has a similar behavior with that in the oversaturated water. The linear relationship of bubble volume versus time (Figure 5c) and the $1/3$ effective power scaling law of $R(t)$ (Figure 5d) are both observed again.

The results for partially degassed water are displayed in Figure 5e–l. In the partially degassed water with $c_\infty/c_s = 0.85$ (Figure 5e,f), we again observe a similar bubble growth dynamics as in the overstaturated water and air-equilibrated water, that is, a linear relationship of bubble volume versus time (Figure 5e) and an effective $1/3$ power scaling law of $R(t)$ (Figure 5f). However, when the gas concentration in partially degassed water is even lower, the growth dynamics is distinctly different. Figure 5g shows the bubble volume versus time in water with $c_\infty/c_s = 0.75$. In this case, the volume no longer linearly increases with $t$. The radius scales as $R(t) \propto t^\alpha$, with an effective exponent $\alpha \approx 0.27$ (Figure 5h). For even lower gas concentrations, namely, $c_\infty/c_s = 0.65$ and $0.55$, $\alpha$ reduces to $0.22$ and $0.07$, respectively.

In plasmonic bubble-related applications, the maximal bubble size (volume) is an important parameter. Figure 5a,c,e indicates that, besides laser power, the growth rate of the bubble volume $\kappa = dV/dt$ is related to the gas concentration. Figure 6 shows the growth rate $\kappa$ as a function of the laser power $P_l$ for the three different gas concentrations of $c_\infty/c_s = 1.2, 0.95,$ and $0.85$. Since $V$ is not proportional to $t$ when $c_\infty/c_s$ is lower than $0.75$, we only extracted $\kappa$ for these three cases. We can see that $\kappa$ increases roughly linearly with the laser power for all gas concentrations, $\kappa = k_{P_l} \cdot P_l$. We find that, for higher gas concentrations, the slope $k_{P_l}$ increases, as seen in the inset of Figure 6.

The above results quantitatively demonstrate the importance of the dissolved gas on bubble growth. Previously, it was shown that, for gas saturation $c_\infty/c_s \approx 1$, the long-term bubble growth (phase 4) is dominated by the influx of gas, which is locally produced around the plasmonic nanoparticles due to heating[17] as the solubility $c_s$ decreases with increasing temperature. All the expelled gas by oversaturation is taken up by the bubble. For such a production-limited growth process, the growth rate is constant, $dV/dt = \kappa$, and consequently, $V(t) \propto t$ or $R(t) \propto t^{1/3}$. Considering the heating transfer originated from gold

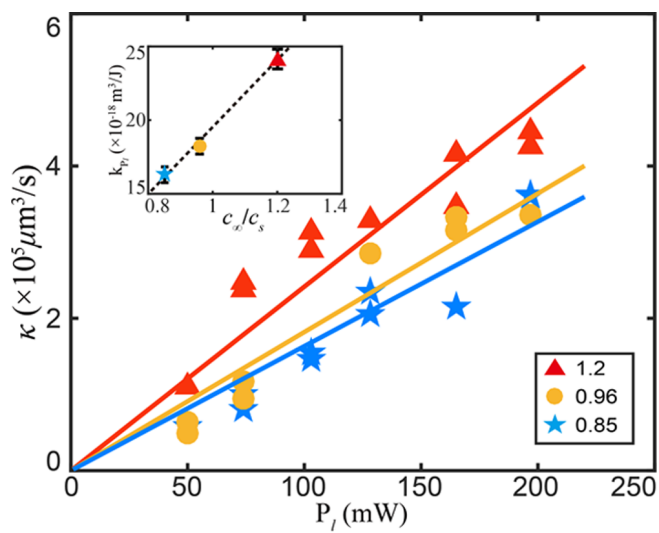

**Figure 6.** Bubble volume growth rates $\kappa$ in the relation $V = \kappa t$ as a function of laser power $P_l$ for different gas concentrations (from $c_\infty/c_s = 0.85$ to $1.2$). $\kappa$ linearly increases with increasing laser power $\kappa = k_{P_l} \cdot P_l$. The prefactor $k_{P_l}$ is found to linearly increase with gas concentration levels $c_\infty/c_s, k_{P_l} \propto c_\infty/c_s$, as shown in the inset figure.

nanoparticles and mass influx, the bubble volume growth rate $\kappa$ is given by[17]

$$\kappa = -\frac{R_g T}{M_g \left(P_\infty + \frac{2\sigma}{R}\right)} \frac{c_\infty}{c_s} \frac{dc_s}{dT} \frac{\beta P_l}{C_w \rho_w} = k_{P_l} \cdot P_l \tag{1}$$

where $M_g$ is the molecular mass of the gas, $R_g$ is the gas constant, $P_\infty$ is the ambient pressure, $\sigma$ is the surface tension, $\rho_w$ is the water density, $C_w$ is the specific heat capacity of water, and $dT$ is the increase in water temperature. Equation 1 shows that, for large $R \gg 2\sigma/P_\infty$, $\kappa$ is proportional to both the relative gas concentration $c_\infty/c_s$ and laser power $P_l$. Again, these linear dependences ($\kappa \propto P_l$ and $k_{P_l} \propto c_\infty/c_s$) are consistent with the results shown in Figure 6.

The effective power law exponent $\alpha$ in the time dependence $R(t) \propto t^\alpha$ of the bubble radius $R(t)$ is used to further investigate the role of the dissolved gas in the bubble growth dynamics (Figure 7). As shown in Figure 7a, the effective exponent $\alpha$ is $\sim 1/3$ for all laser powers if $c_\infty/c_s$ exceeds $0.85$. For lower gas concentrations, $c_\infty/c_s = 0.55$–$0.75$, the effective exponents are smaller than $1/3$ and slightly increase with increasing laser power $P_l$. These results are consistent with our previous findings, where $R(t) \propto t^{1/3}$ and $R(t) \propto t^{0.07}$ for experiments in air-equilibrated water and degassed water, respectively.[17] Figure 7b shows the effective exponent as a function of $c_\infty/c_s$. As seen before, for a given laser power, the exponent increases with increasing gas concentration. The dependence of $\alpha$ on both $c_\infty/c_s$ and $P_l$ is summarized in the three-dimensional plot in Figure 7c.

The above results reveal how the dissolved gas controls the long-term bubble growth dynamics. For large gas concentrations, there is a constant influx of gas into the bubble, leading to a linear growth of bubble volume. The transfer of heat per unit area to the bubble/water interface becomes so small that there is insufficient energy available to overcome the large latent heat of the vaporization barrier. As a result, gradually, the bubble is thermally decoupled from the nanoparticles. The amount of vapor molecules inside the bubble stabilizes. Thus, the bubble growth will be dominated by expelled gas due to the local gas oversaturation.

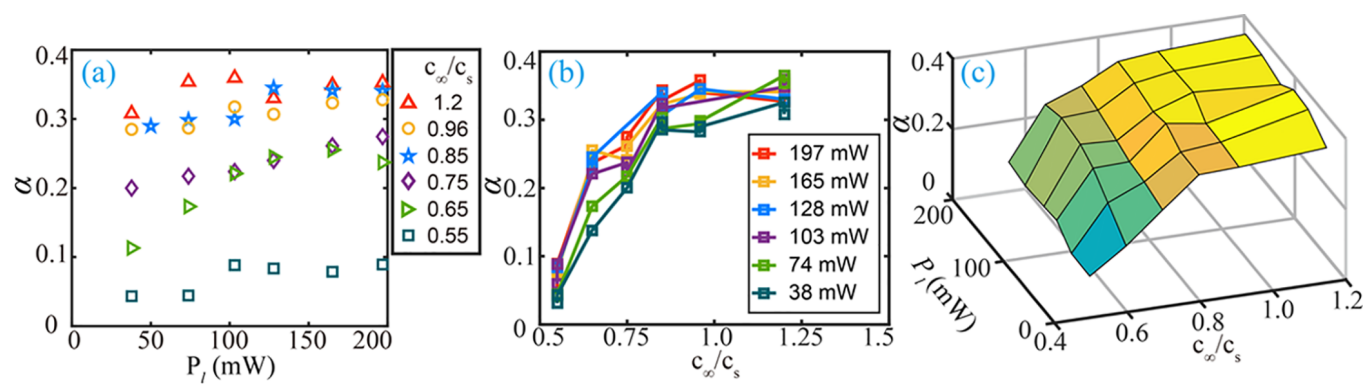

**Figure 7.** Effective power law exponent $\alpha$ in $R(t) \propto t^{\alpha}$ as a function of laser power $P_l$ and gas concentration $c_{\infty}/c_s$. (a) Effective exponent $\alpha$ as a function of laser power for six different gas concentrations $c_{\infty}/c_s$. (b) Effective exponent $\alpha$ as a function of gas concentration for different laser powers $P_l$. (c) 3D plot of $\alpha$ as a function of $c_{\infty}/c_s$ and $P_l$. Approximately, $\alpha$ monotonously increases with both $P_l$ and $c_{\infty}/c_s$.

For large enough initial gas concentration $c_{\infty}/c_s \geq 0.85$, the effective exponent $\alpha$ in the dependence $R(t) \propto t^{\alpha}$ saturates at an exponent of 1/3, which is the limit of value for the diffusive growth. In addition, for strongly undersaturated water, the exponent is smaller than 1/3. The latter can easily be understood since the solubility of gas in water decreases with increasing temperature. This means that the water first has to be heated up to a temperature where it becomes supersaturated. Once the water has reached this supersaturated regime, the bubble grows by gas that is expelled from the supersaturated water. With increasing bubble size, the thermal energy transferred per unit area of the bubble/water interface will rapidly decrease. As a result, the amount of oversaturated gas per unit volume near the bubble/water interface will decrease also. At some point, the energy transferred to the bubble/water interface becomes so low that no further gas is expelled and the bubble growth terminates.

Notably, at higher laser powers, the effective scaling exponent $\alpha$ for $c_{\infty}/c_s = 0.65$ and 0.75 in Figure 7a becomes approximately equal. The reason is that the relative oversaturation of gas, which dominates the long-term growth of plasmonic bubble, depends on the original amount of dissolved gas (at room temperature) as well as on the temperature (and thus, the laser power). If the laser power is larger, then the water will be heated up to higher temperatures, leading to a larger gas oversaturation (the solubility of air in water decreases with increasing temperature). As a result, the effect of the initial gas concentration will become weaker at higher laser power. We found a similar behavior for $c_{\infty}/c_s = 0.85$, 0.96, and 1.2. The effective scaling exponent $\alpha$ for these three different but relatively large gas concentrations is very comparable at higher laser powers.

## CONCLUSIONS

In summary, the effect of the dissolved gas on the dynamics of plasmonic bubble nucleation in the early phase and in the long-term growth regime has been systematically studied. In the early phase, lower gas concentrations lead to a longer delay time $\tau_d$, larger maximum volume, and higher nucleation temperature, which indicate that the dissolved gas facilitates bubble nucleation. We have found a linear relation between the bubble volume and the total energy. The prefactor of this linear relation is the same for all gas concentrations, reflecting that the bubbles that form in the first stage of the irradiation process are vapor bubbles. Regarding the long-term growth dynamics, we have shown that the growth rate $\kappa$ of the bubble volume monotonically increases with gas concentration $c_{\infty}/c_s$ and laser power $P_l$. Moreover, the experimental results show a linear dependence of $\kappa = k_{P_l}P_l$, where $k_{P_l}$ linearly increases with $c_{\infty}/c_s$. The radius $R(t)$ of the plasmonic bubbles follows the power law dependence $R(t) \propto t^{\alpha}$. For all laser powers, the effective exponent $\alpha$ for all laser powers is ~1/3 when $c_{\infty}/c_s$ is larger than 0.85. However, for lower $c_{\infty}/c_s$ values, the effective exponent is smaller than 1/3 and monotonously decreases with decreasing $P_l$ and $c_{\infty}/c_s$. For strongly degassed water, the exponent $\alpha$ is smaller than 1/3 because the water first has to be heated up to a temperature where it becomes supersaturated and gas can be expelled.

## AUTHOR INFORMATION

**Corresponding Authors**
*E-mail: wangyuliang@buaa.edu.cn (Y.W.).
*E-mail: d.lohse@utwente.nl (D.L.).

**ORCID**

Xiaolai Li: 0000-0002-4912-2522
Yuliang Wang: 0000-0001-6130-4321
Hai Le The: 0000-0002-3153-2937
Xuehua Zhang: 0000-0001-6093-5324
Detlef Lohse: 0000-0003-4138-2255

**Notes**
The authors declare no competing financial interest.

## ACKNOWLEDGMENTS

The authors thank the Dutch Organization for Research (NWO) and the Netherlands Center for Multiscale Catalytic Energy Conversion (MCEC) for the financial support. D.L. acknowledges the financial support by an ERC Advanced Grant (no. 740479) and by NWO-CW. Y.W. acknowledges the financial support by an NWO travel grant, the National Natural Science Foundation of China (grant no. 51775028), Beijing Natural Science Foundation (grant no. 3182022), and ERC-NSFC joint program (grant no. 11811530633). X.Z. acknowledges the support of the Natural Sciences and Engineering Research Council of Canada (NSERC) and Future Energy Systems (Canada First Research Excellence Fund).